\documentclass[aps,pra,twocolumn,showpacs,superscriptaddress]{revtex4-2}

\usepackage[dvips]{graphicx}
\usepackage{url}
\usepackage{bm}
\usepackage{amsmath,amssymb}
\usepackage{color,soul}
\usepackage{lineno}
\usepackage{xcolor}
\usepackage{lipsum}
\usepackage{booktabs}

\usepackage[unicode]{hyperref}

\hypersetup{pdfstartview=FitV,
citecolor=blue,
colorlinks=true,
linkcolor=blue}

\newcommand{\vA}{\mathbf{A}}

\newcommand{\vd}{\mathbf{d}}

\newcommand{\vp}{\mathbf{p}}

\newcommand{\vvr}{\mathbf{r}}
\newcommand{\vcf}{\mathbf{F}}

\newcommand{\he}{\mathbf{\hat e}}

\graphicspath{{figs/}}

\begin{document}

\title{Directional loss of contrast by dephasing in temporal double-slit interferometry}

\author{M. A. H. B. Md Yusoff}
\author{H. B. Ambalampitiya}
\author{J.M. Ngoko Djiokap}
\affiliation{Department of Physics and Astronomy, University of Nebraska,
  Lincoln, Nebraska 68588-0299, USA}

\date{\today}

\begin{abstract}
\textcolor{black}{Attosecond streaking camera is an ex situ technique in which a linearly polarized (LP) XUV attopulse produces an electron wavepacket by photoionization in the presence of an IR femtopulse. By moving the two synchronous oppositely circularly polarized XUV pulses (that make the ionizing LP pulse) apart in time, we propose an attosecond double-slit streak camera scheme to see information loss in polarization-dependent two-slit phenomena. Such streaking interferogram is then composed of several Feynman's thought experiments in time domain, in which electrons are affected by the IR pulse as they exit the two attoslits upon XUV ionization processes to form Archimedean spiral patterns. As a first proof-of-principle, when the IR femtofield and first XUV attopulse are synchronous, a loss of contrast through dephasing is seen in the simulated momentum and energy distributions of the photoelectron. It is shown that the contrast between bright and dark interference fringes diminishes predominantly along the IR-field polarization axis and is sensitive to the IR-field waveform. Our which-time information scheme provides further confirmation of the wave-particle duality.} 
\end{abstract}

\maketitle


One of the most groundbreaking experiments in optics and quantum mechanics is the space two-slit scheme first realized by Young using light~\cite{Young1801}. It relies on interference of matter waves: a key concept of quantum theory, which has been confirmed experimentally by electron diffraction \cite{Davisson1927,Thomson1927}. Its scope was greatly expanded not only by the famous ``thought experiment'' for electrons proposed in 1965 by Richard Feynman~\cite{Feynman1965,Cohen-Tannoudji1977}, but also by recent studies on the loss of contrast~\cite{Rui-Feng2014,Chen2018} in space two-slit interferometry, and finally by a variety of two-slit schemes in time domain~\cite{Moshinsky1952,Szriftgiser1996,Wollenhaupt2002,Lindner2005,Ngoko2015,Pengel2017,Cheng2020,Keramati2020,Xiao2019,Ngoko2021,Tirole2023,Harin2024}.  The basic idea in the Feynman’s thought experiment is to place a strong light source behind the wall and between the two holes, with the ultimate goal of ``watching'' the electrons as they exit the spaced two slits since electric charges scatter light~\cite{Feynman1965}. \textcolor{black}{All these two-slit interference studies have been instrumental in fueling fundamental quantum-mechanical research (see~Refs~[1-5] in~\cite{He2020}). They also drove numerous practical applications, including the detection of orbital symmetries~\cite{Harin2024} and structure~\cite{Kushawaha2013} in molecules, measurement of light coherence~\cite{Lyubomirski2016}, and retrieval of nuclei motion~\cite{Canton2011}.}  

In optical and matter-wave interferometry, a loss of contrast is a limit on the detection capability of such devices~\cite{Chen2018}. The studies of loss of contrast restricted in spatial double-slit interferometry revealed that it can be ascribed either to dephasing or decoherence processes~\cite{Rui-Feng2014,Chen2018}. A key difference between these two processes~\cite{Chen2018} is that dephasing is time reversible with constant entropy $S$, whereas decoherence is time irreversible with \textcolor{black}{varying} $S$~\cite{Nielsen2011}. 
Inspired by previous studies on loss of contrast in space two-slit interferometry, \textcolor{black}{what remains to be accomplished is its study in time domain. To bring out another facet of the wave-particle duality, is it possible to use tailored ultrashort ionizing pulses to realize a Feynman's thought experiment on an attosecond timescale? }


In the time two-slit scheme~\cite{Wollenhaupt2002} where electrons are produced by photoionization by a pair of time-delayed, counter-rotating circularly polarized (CRCP), XUV laser attopulses, Archimedean spiral patterns were discovered~\cite{Ngoko2015} by Ramsey interference~\cite{Ramsey1950} in the in-plane photoelectron momentum distribution (PMD). Confirmed experimentally numerously (see, e.g.,~\cite{Pengel2017}) using tailored femtosecond pulse shapes, they have opened up an interdisciplinary field across different branches of physics (see, e.g., Refs.~[9-46] in~\cite{Harin2024}). As the electron-matter waves exit the two slits and interfere to form spiral patterns in the PMD, can those electrons be affected by a synchronized intense light source? \textcolor{black}{Given the contrast in the timescales between the ionizing and streak pulses, whether the light source can interact with the electron exiting the first or second slit, or both, is nontrivial.} 


\begin{figure}[b]
  \centering
  \includegraphics[width=0.9\linewidth]{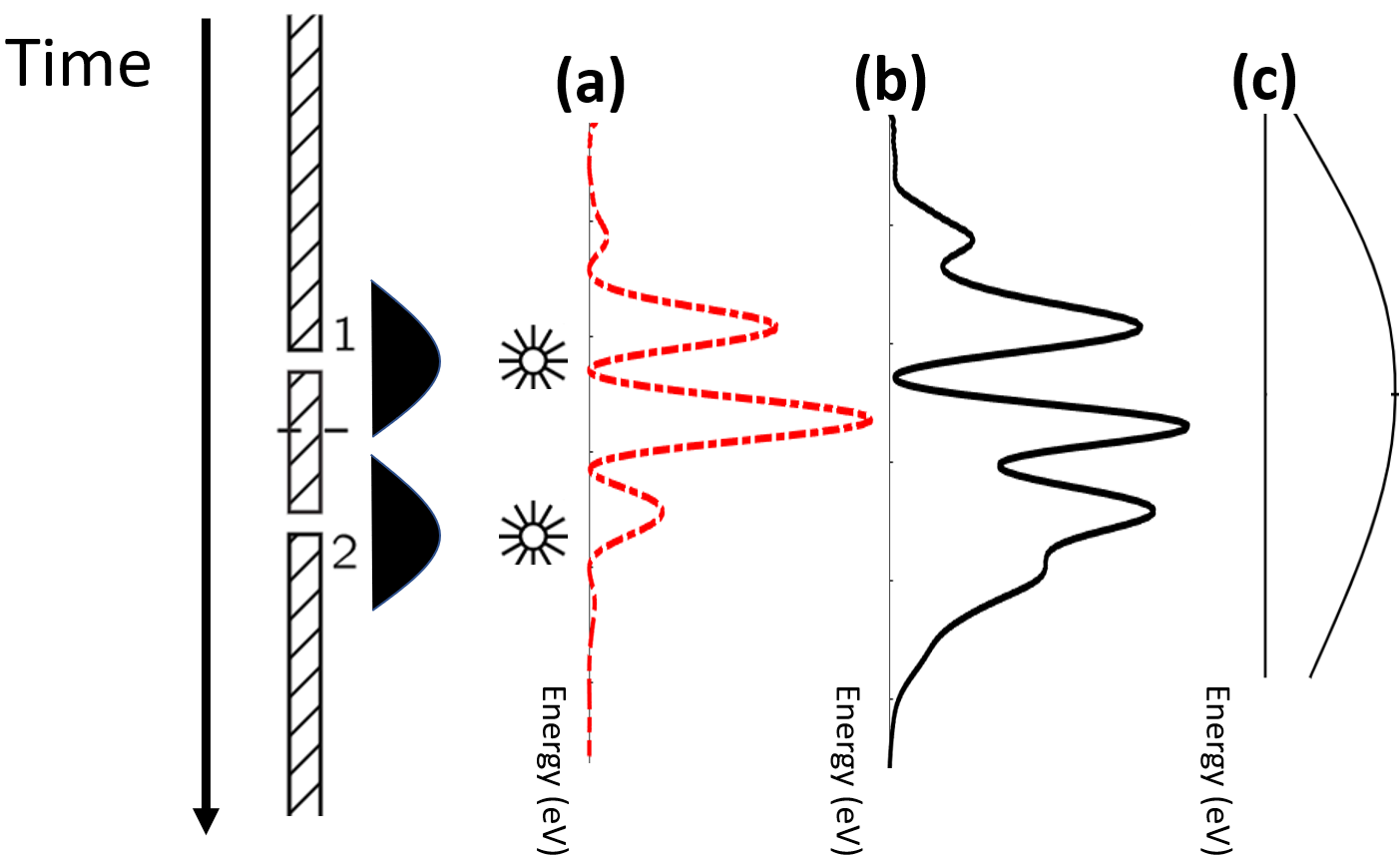}
  \caption{\textbf{A Feynman thought experiment scheme on an attosecond timescale.} One or two noninvasive IR light is or are placed at the locations of the two slits, where a pair of time-delayed XUV pulses create a pair of electrons (see the shaded lumps) upon ionization. (a)~Excellent contrast between bright and dark interference fringes is seen in the PED in the absence of IR fields. Partial~(b) and total~(c) loss of contrast are depicted when the IR fields are turned on.}
\label{Fig1}
\vspace{-0.65cm}
\end{figure}

In this Letter, \textcolor{black}{by synchronizing the light source with respect to the first XUV pulse} we realize the Feynman’s thought experiment on an attosecond timescale by placing one or two IR femtolight at the location of the attoslits separated roughly by $1$~fs, see Fig.~\ref{Fig1}. We vary the waveforms and intensities of the single-cycle IR fields and monitor the change in the contrast between bright and dark interference fringes from two observables: the PMD and the photoelectron energy distribution (PED). In the absence of IR field, excellent contrast is achieved, as depicted in Fig.~\ref{Fig1}(a) for the PED. When one IR light intense enough is placed at the first attoslit, then manifests a CEP-dependent loss of contrast along the IR-field polarization axis, since the electron exiting either the first or second slit can be affected by the IR field depending on its waveform. \textcolor{black}{In attoscience, this scheme for synchronous CRCP XUV pulses reduces to a single point of the attosecond streak camera: a powerful tool used for the first characterization of isolated attopulses~\cite{Hentschel2001}, and for attosecond chronoscopy of photoemission in matter~\cite{Pazourek2015}.}  Turning on another IR light at the other attoslit leads to optical interference with the first IR field, where electrons exiting the two slits are affected by the combined light. This results in a significant CEP-sensitive loss of contrast, as depicted in Fig.~\ref{Fig1}(b) for a comparison with Fig.~\ref{Fig1}(c) for a total loss of contrast. These losses of contrast are directional as the contrast remains excellent in the direction perpendicular to the IR-field polarization. \textcolor{black}{In matter-wave interferometry, such loss of contrast can be used to detect the existence of stray radiation in the double-slit experimental setup.} 

A key requirement for observing significant loss of contrast is that near the peak of an attosecond slit, the vector potential $\vA_{\text{IR}}(t)=(\sqrt{I_0}/\omega)f(t)\sin(\omega t+\phi) \he_x$ of the linearly-polarized light must be weak enough not to ionize matter but strong enough to impart substantial momentum shift to photoelectrons liberated by the XUV pulses~\cite{Nisoli2017}. While this can be achieved by using infrared or terahertz light source because of their small value of carrier frequency $\omega$, the IR fields are preferential in light of future attosecond streaking experiments that manipulate the electron spiral phenomenon, which requires broad XUV pulse bandwidth to resolve interference fringes. In our proposed polarization-scheme in Fig.~\ref{Fig1} to study the angularly-resolved loss of contrast with a focus on the dephasing processes, the particles incident to the temporal two slits are XUV CRCP ``photons''; however, the particles leaving the slits upon ionization are photoelectrons with energy $E=p^2/2$. Without any loss of generality, here we consider the illustrative case of He atom, single ionized by one-photon absorption. Each XUV gaussian-pulse has two cycles; an intensity $I_{0,\text{XUV}}=0.1$~PW/cm$^2$; duration (FWHM) $\tau_{0,\text{XUV}}=229.8$~as; and carrier frequency $\omega_{\text{XUV}}=36$~eV greater than $E_b=24.59$~eV, the binding energy for the $^1$S$^e$ ground state of He. To interact with these electron waves as they exit the attoslits, we use the shortest (single-cycle) $800$~nm IR gaussian-field ($\omega=1.55$~eV), having a tunable intensity $I_0$ and stable CEP $\phi\equiv\phi_{\text{IR}}$. In contrast to~\cite{Bertolino2020}, having an XUV pulse duration much shorter than the IR period of $2.7$~fs implies that discrete peaks corresponding to interaction with $q$ photons in the continuum are hidden under the broader photoelectron spectrum created by the XUV pulse.

For these field parameters, we deploy the RMT code~\cite{Moore2011,Brown2020,Lysaght2009,Clarke2018} within the electric dipole approximation of the semi-classical theory of laser-matter interaction to describe this process. By solving the full-dimensional time-dependent Schr\"{o}dinger equation (TDSE) in which the classical description of the laser field and quantum treatment of the quantum system are valid, any loss of contrast reported here can only proceed by dephasing, not by decoherence. \textcolor{black}{In laser-matter interaction, decoherence may manifest when there is quantum entanglement between the system and light. This may be facilitated by the occurrence of quantum fluctuations in either the system (e.g., in the ground state~\cite{Richard2025}) or the laser fields (as done in the emerging field of attosecond quantum optics, see, e.g., Refs.~\cite{Yi2025,Bhattacharya2023}). This is not the case here as within the laser approximation~\cite{Joachain1994} the intense laser fields carry quadrillions of photons in a coherent state.}

\begin{figure}[t]
  \centering
  \includegraphics[width=0.49\linewidth]{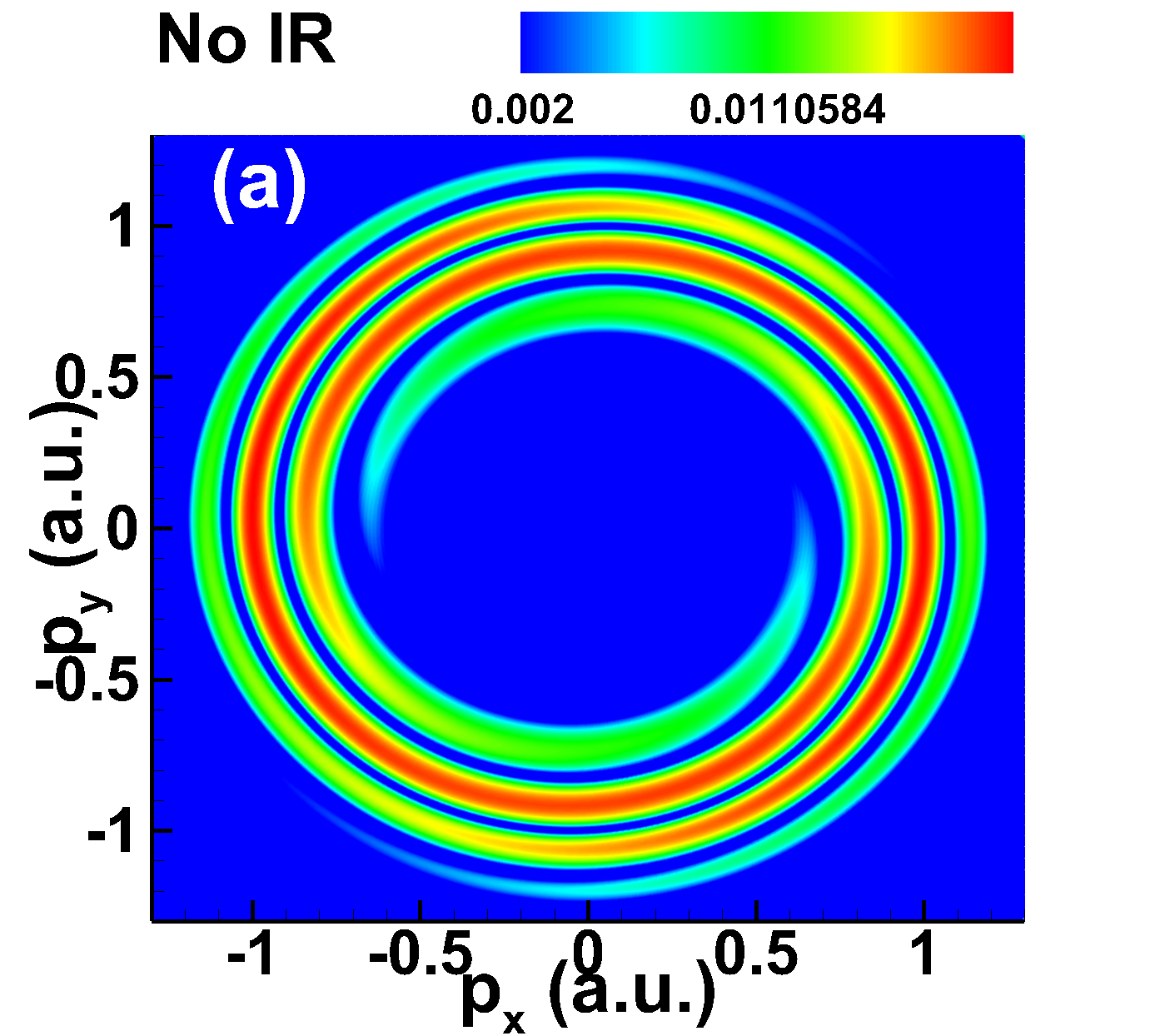}
  \includegraphics[width=0.49\linewidth]{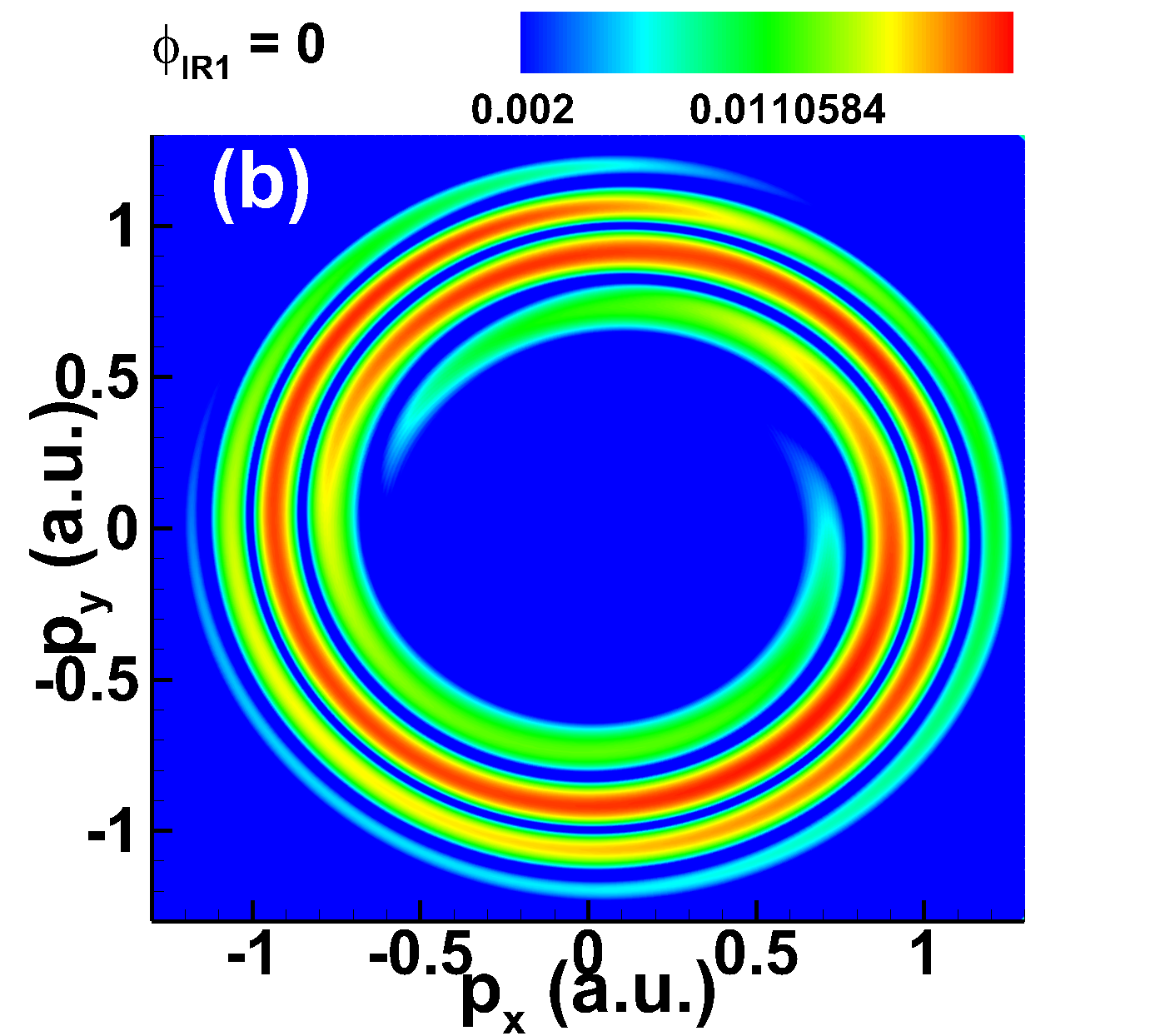}\\
  \includegraphics[width=0.49\linewidth]{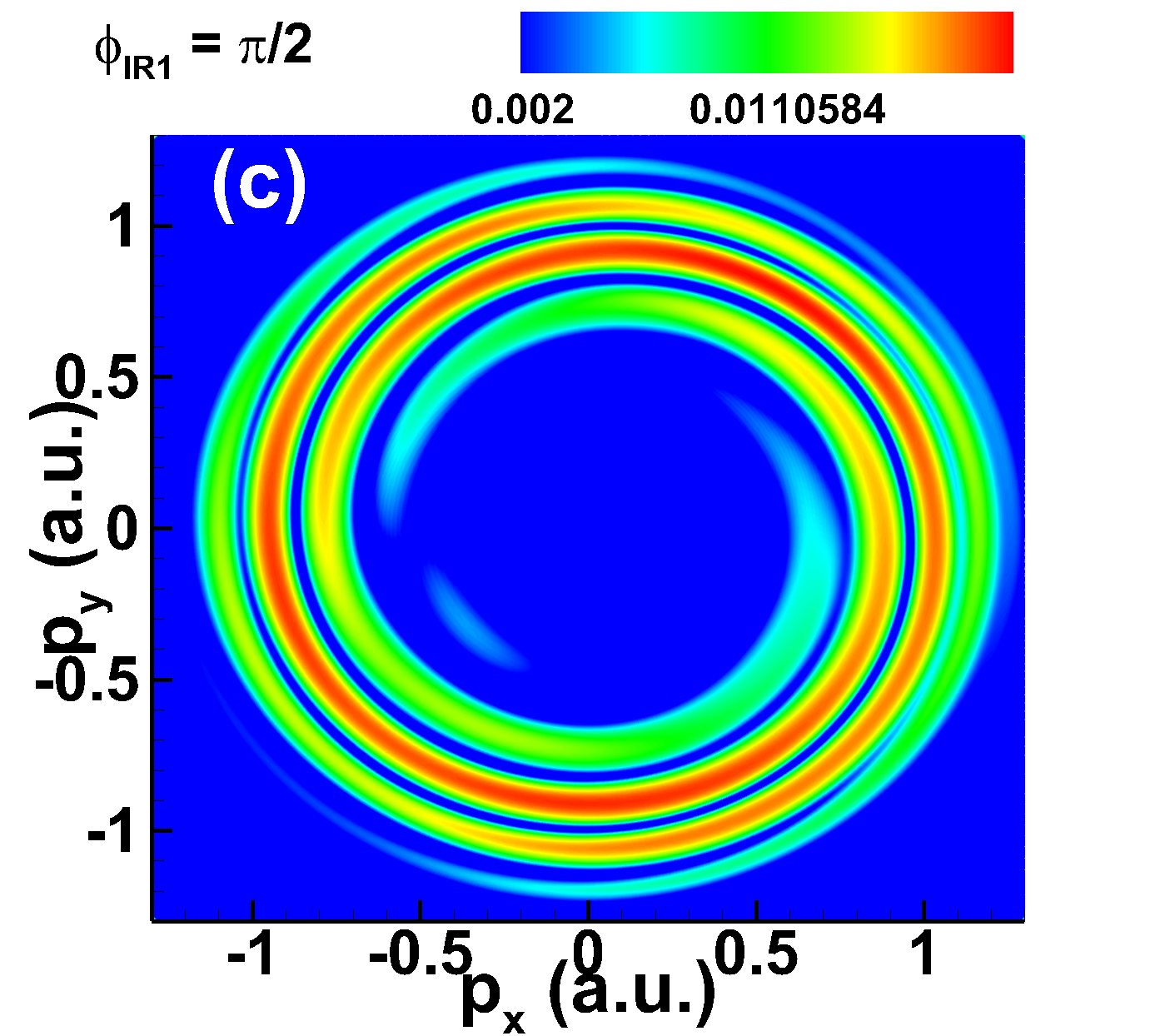}
  \includegraphics[width=0.49\linewidth]{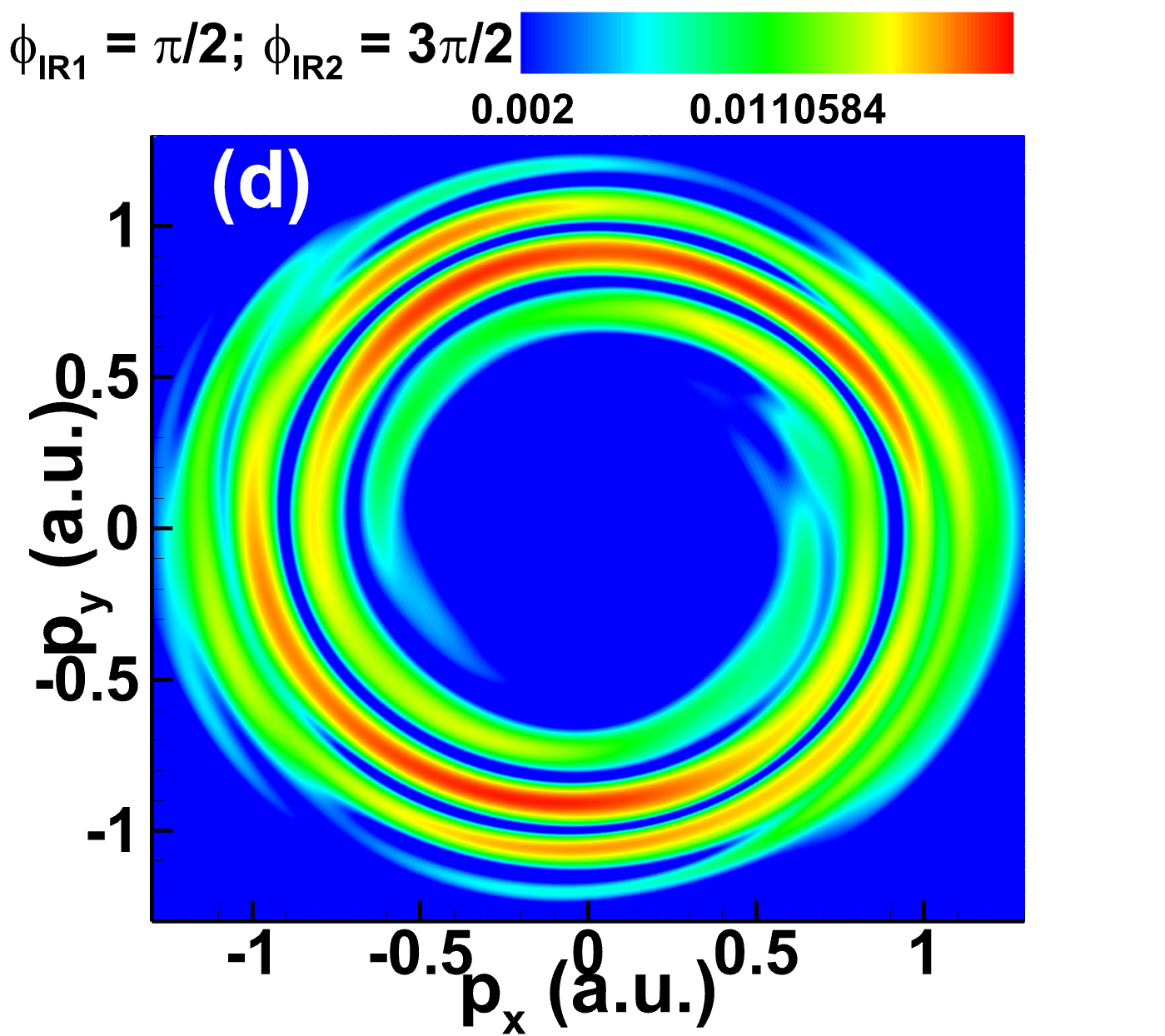}
  \caption{\textbf{Effects of infrared light on electrons as they exit the temporal two slits to form Archimedean spirals through Ramsey interference:} PMDs in the polarization plane for photoionization of He atoms produced in the absence (a) or presence of one [(b)-(c)] and two~(d) IR fields. The single-cycle gaussian IR field has a wavelength $\lambda_{\text{IR}}=800$~nm, an intensity $I_{0,\text{IR}}=1$~TW/cm$^2$, and various CEP values: (b)~$\phi_{\text{IR}1}=0$, (c)~$\phi_{\text{IR}1}=\pi/2$, and (d)~$\phi_{\text{IR}1}=\pi/2$, $\phi_{\text{IR}2}=3\pi/2$. The time-dependent electric fields of the $x$-component of the XUV pulses and IR fields for various waveform schemes are shown in~\cite{SM}.
  }
\label{Fig2}
\vspace{-0.5cm}
\end{figure}

\textcolor{black}{Our TDSE results for the PMDs produced by time-delayed CRCP XUV laser pulses in the presence or absence of the IR field are shown in Fig.~\ref{Fig2}. The effects of the light source on the PMDs either through the shift with~[Figs.~\ref{Fig2}(c,d)] or without~[Fig.~\ref{Fig2}(b)] raising of dark fringes are tangible.}  Our focus is on threshold electrons whose range $0\leq\Delta E\leq 30$~eV is dictated by the one-photon pulse bandwidth. The time delay between the two XUV pulses is $\tau_{\text{XUV}}=3\tau_{0,\text{XUV}}$, meaning that the created pair of electron wave packets are initially separated in time. In the absence of the IR field, when the electron wave packet is created by the second slit, the one emerging from the first slit has already accumulated a phase: $\Psi\equiv (E+E_b)\tau_{\text{XUV}}+\phi_{12}$, where the first term is the Ramsey phase, and the second term is the CEP difference of the attopulses. Therefore, according to the electric dipole selection rules a pair of essentially $^1$P$^o$ electron wave packets with azimuthal quantum number $M_L=\pm1$ (see the ionization channel populations in Fig.~\ref{Fig3}(a)) spread out due to dispersion and interfere to form a spiral pattern seen in Fig.~\ref{Fig2}(a) in the polarization plane ($\theta=\pi/2$), where an excellent contrast between bright and dark interference fringes is visible. Figure~\ref{Fig3}(a) also shows the negligible first above-threshold ionization~\cite{Agostini1979} channels $(2,-2)$ and $(2,+2)$ produced from the channels $(1,-1)$ and $(1,+1)$, for which the much higher photoelectron energies lie outside the above-mentioned range $\Delta E$. 

An analytical expression for the triply differential probability (TDP), $\mathcal{W}(\vp)\equiv dW/d\vp=|\mathcal{A}|^2$, for photoionization in the absence or presence of the IR field can be derived from the transition amplitude~\cite{Lin2018} obtained within the strong-field approximation (SFA) framework:
\begin{equation}\label{Amplitude}
\mathcal{A}=-i\int_{-\infty}^{+\infty} \vd(\vp+\vA_{\text{IR}})\cdot \vcf_{\text{XUV}}(t-\tau_{\text{IR}})~e^{i[\varepsilon t+\Phi_{\text{IR}}(\vp,\phi,t)]} dt,
\end{equation} 
where $\varepsilon\equiv(E+E_b)$; $\vd(\vp+\vA_{\text{IR}})\approx\vd(\vp)$ is the dipole moment in standard SFA~\cite{Lin2018,Cheng2020}; $\vcf_{\text{XUV}}(t)$ is the electric field of the ionizing pair of CRCP pulses with envelope $f_{\text{XUV}}(t)$; and $\Phi_{\text{IR}}(\vp,\phi,t)=\int_{t}^{+\infty} [\vp\cdot\vA_{\text{IR}}(t')+\vA^2_{\text{IR}}(t')/2]dt'$ is the phase induced by the IR field.

\begin{figure}[t]
  \centering
  \includegraphics[width=0.49\linewidth]{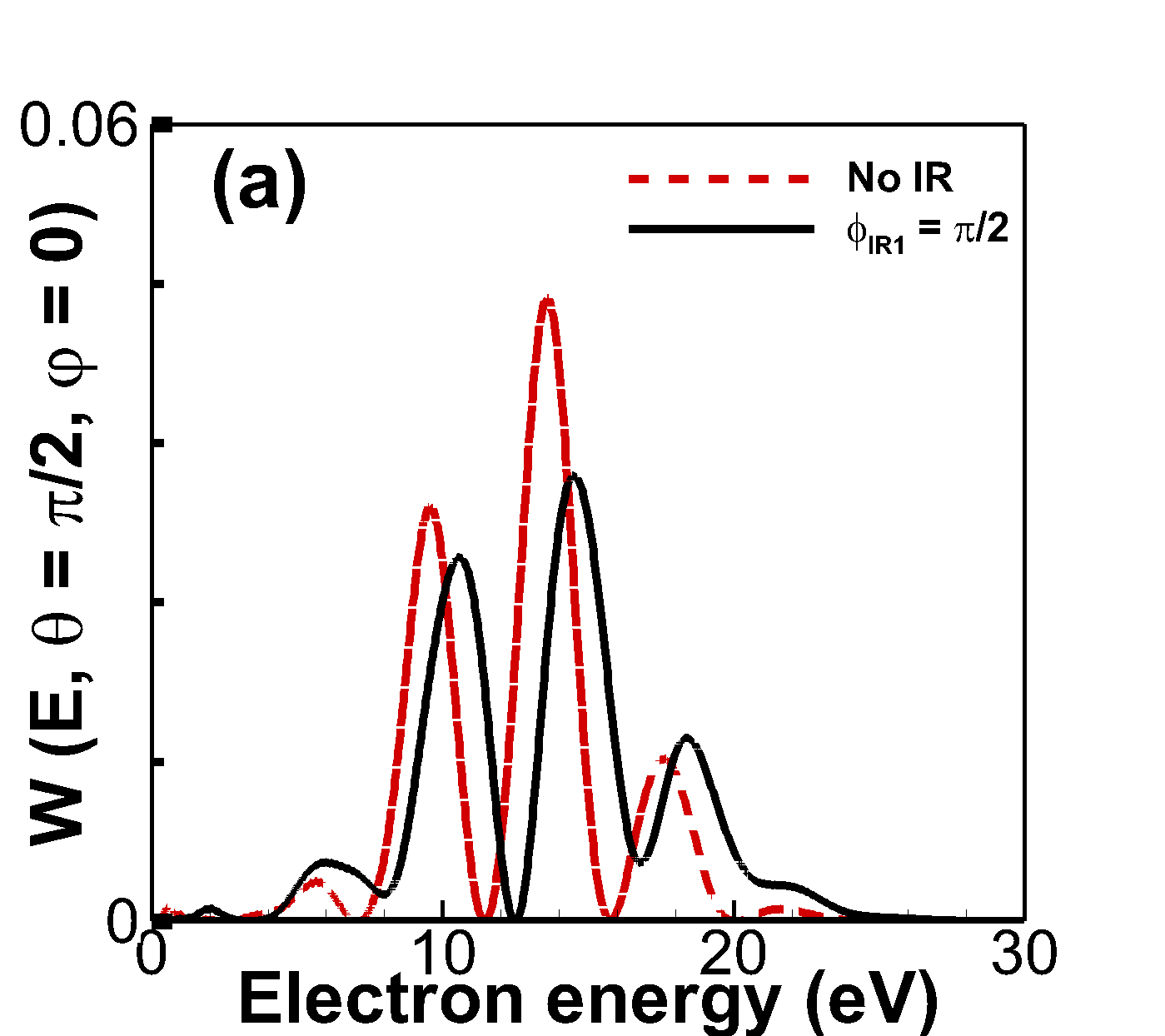}
  \includegraphics[width=0.49\linewidth]{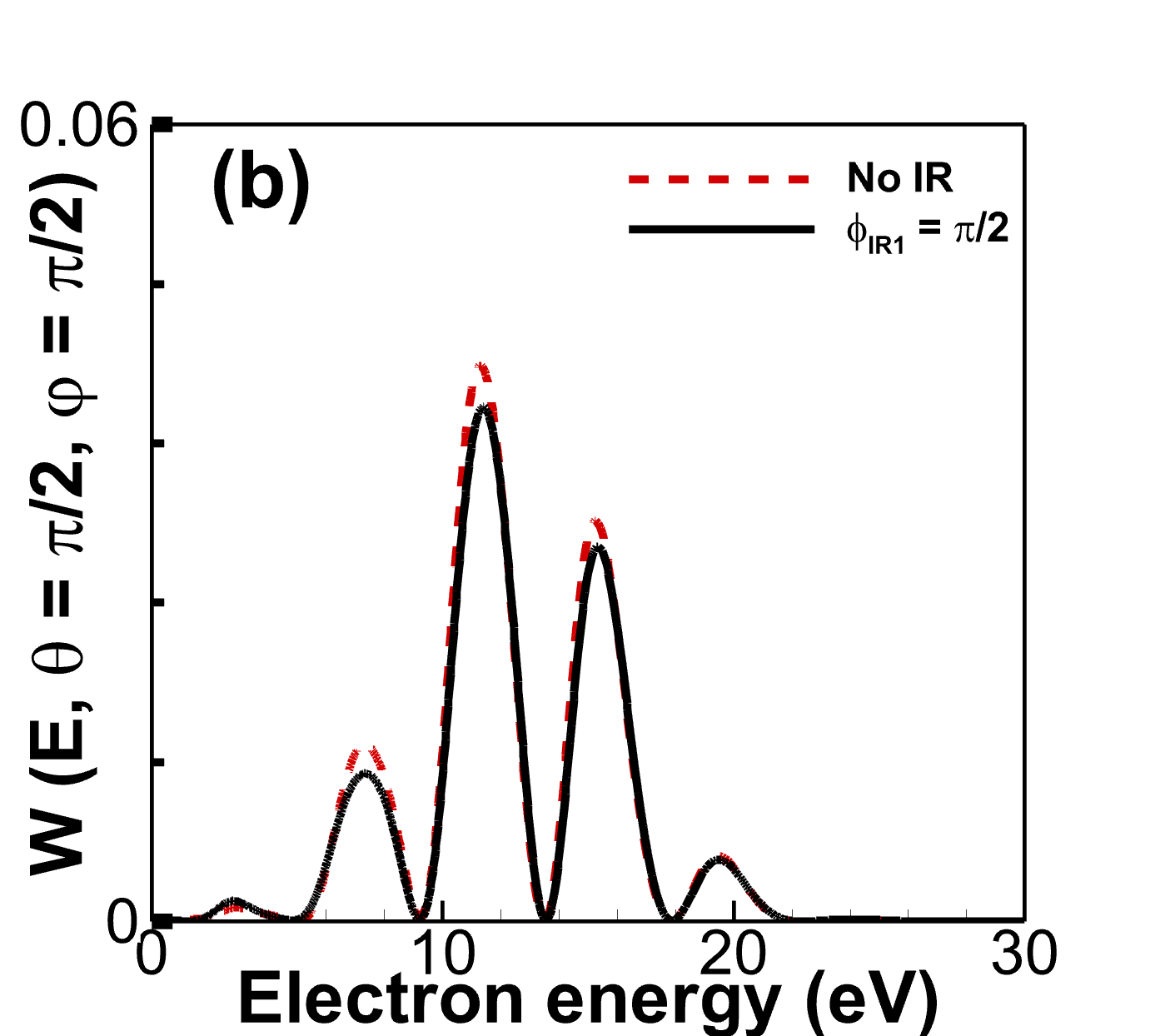}\\
  \includegraphics[width=0.49\linewidth]{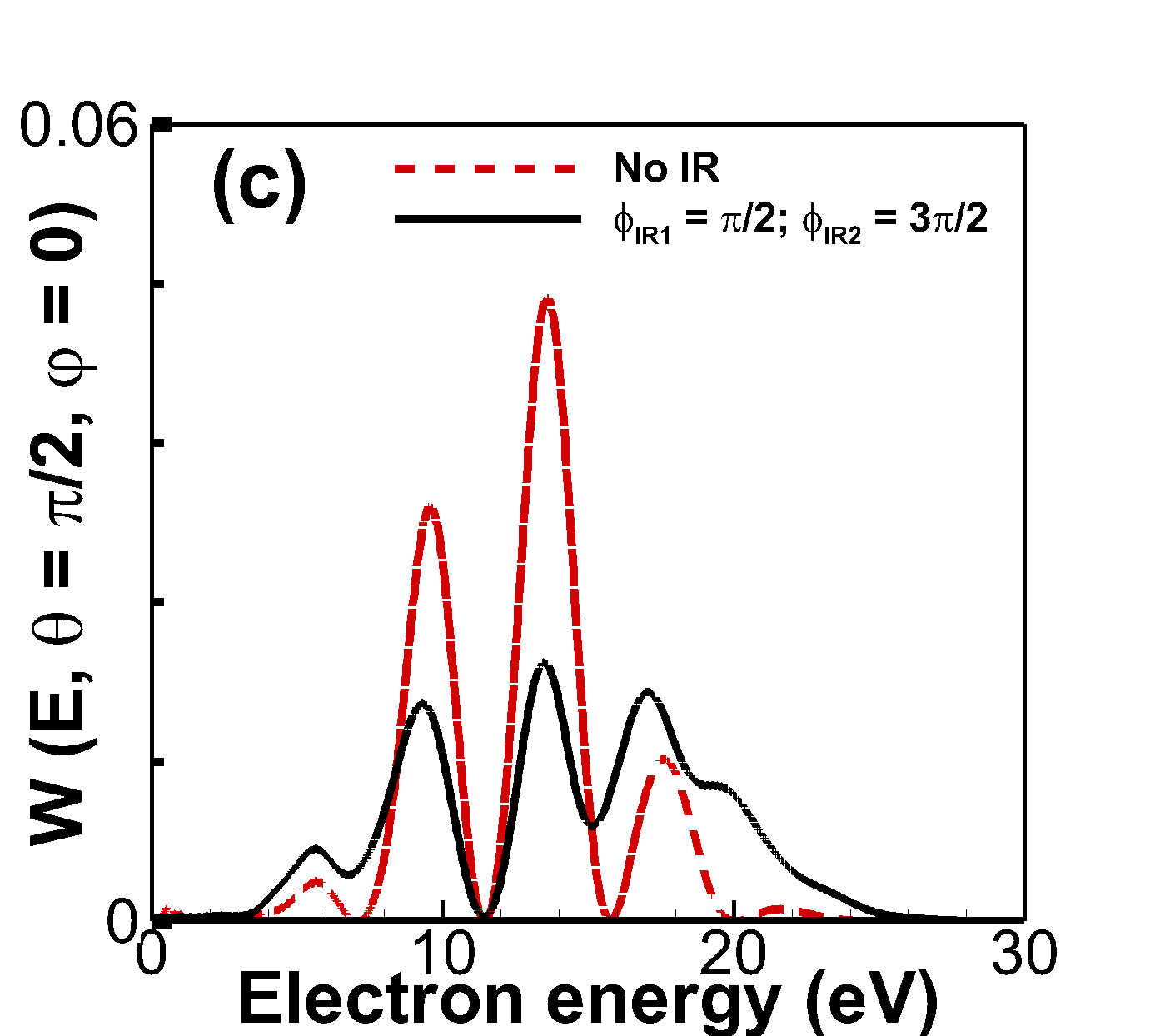}
  \includegraphics[width=0.49\linewidth]{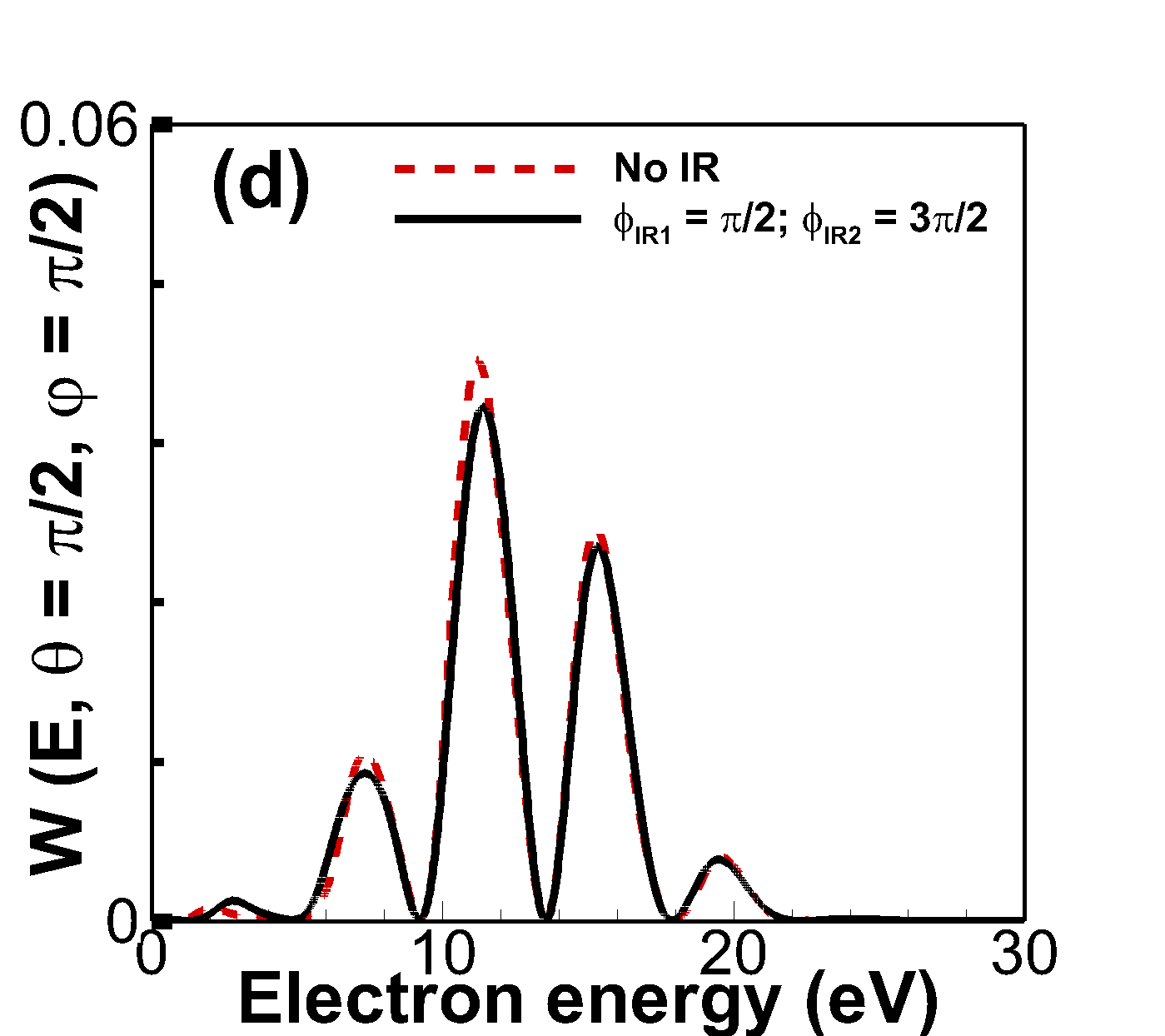} 
  \caption{\textbf{Visualization of the loss of contrast in a PED.} [(a),(c)]~forward emission ($\varphi=0$)  and [(b),(d)]~orthogonal emission ($\varphi=\pi/2$) in the polarization plane ($\theta=0$) in the case of one IR field with a CEP of $\pi/2$ (top) and two IR fields with CEPs of $\pi/2,3\pi/2$ (bottom). In each panel, results for no IR field is shown in red dashed curves for comparison.   }
  \label{Fig4}
  \vspace{-0.5cm}
\end{figure}

\begin{figure}[t]
  \centering
  \includegraphics[width=0.49\linewidth]{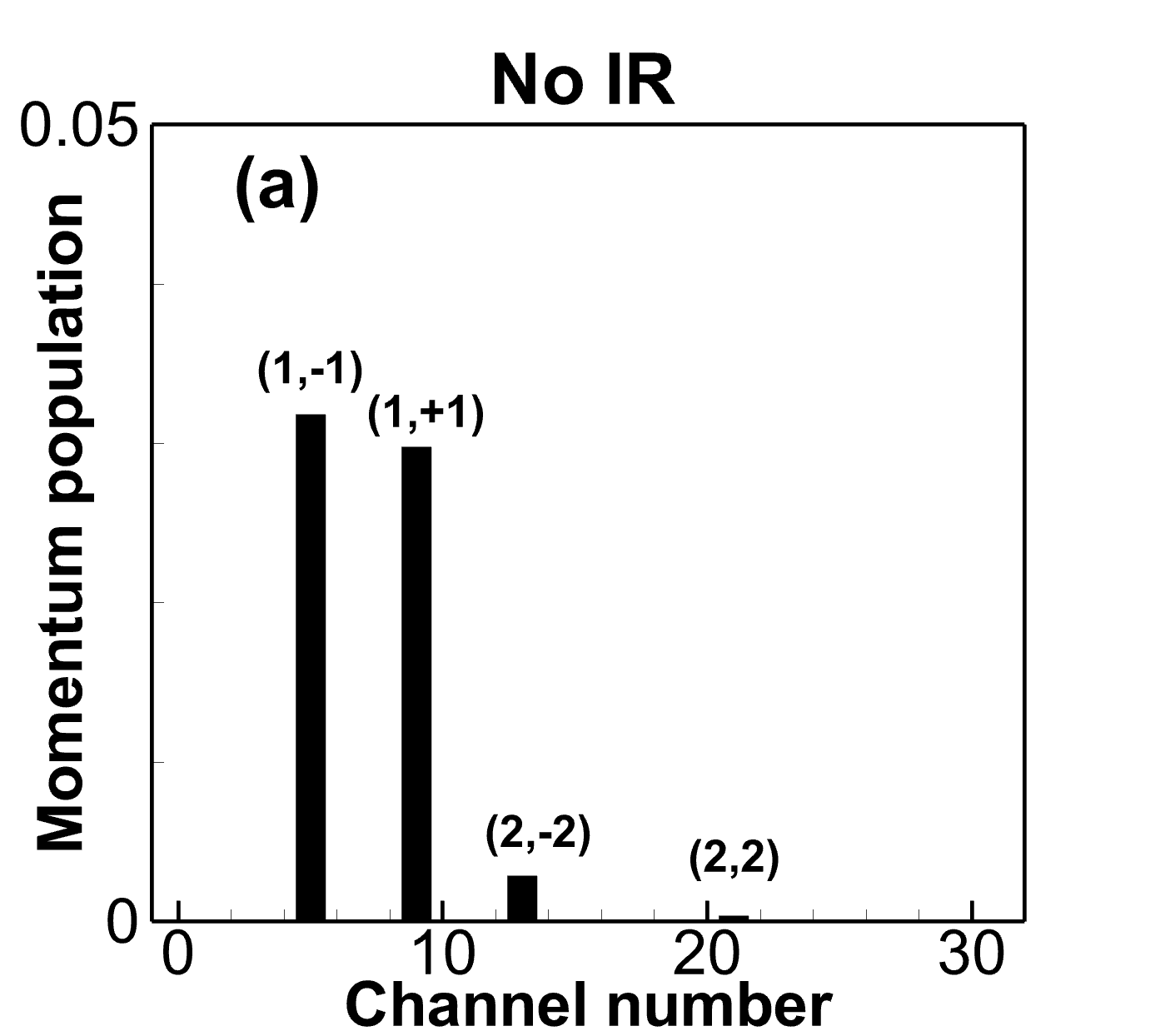}
  \includegraphics[width=0.49\linewidth]{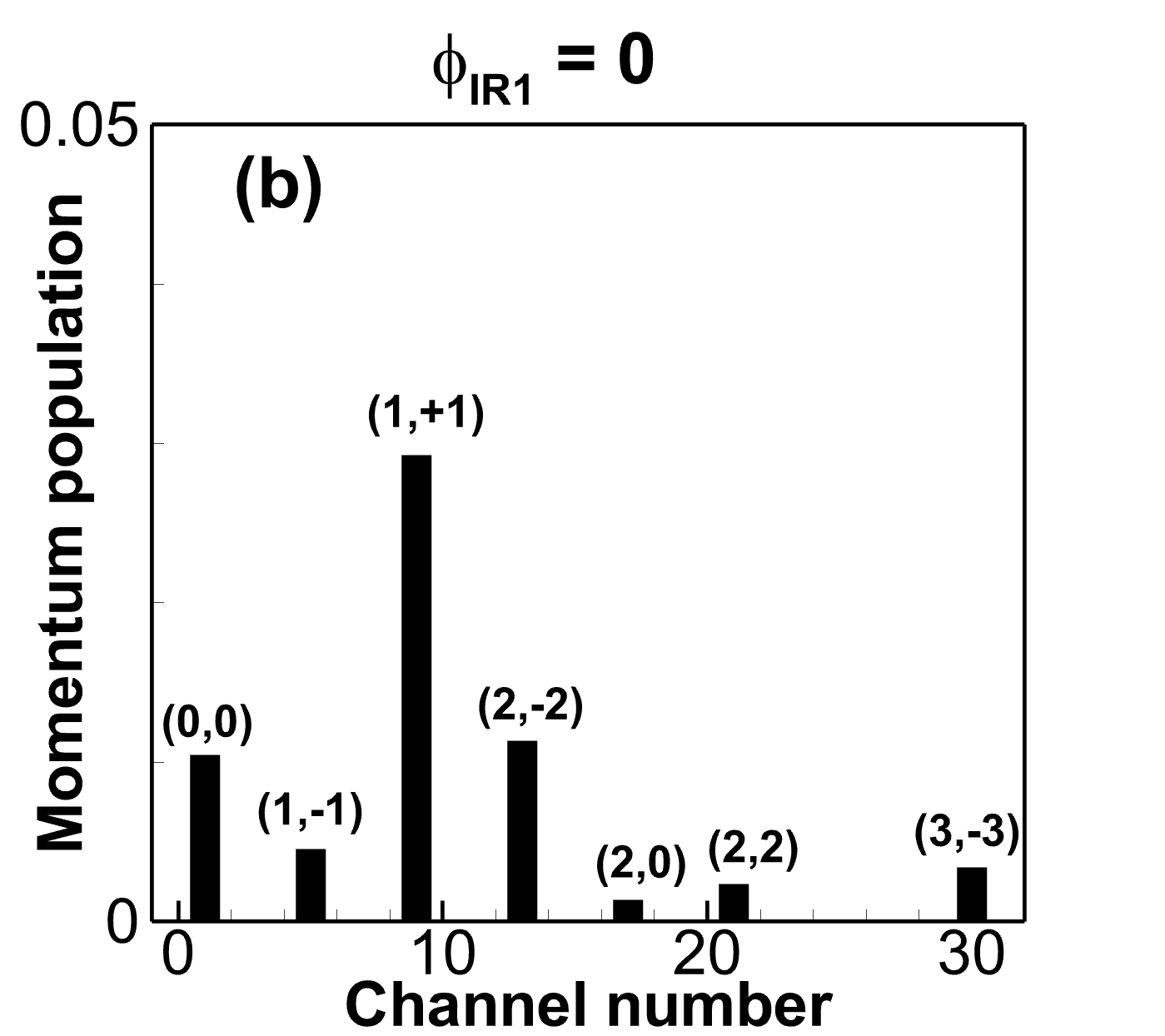}\\
  \includegraphics[width=0.49\linewidth]{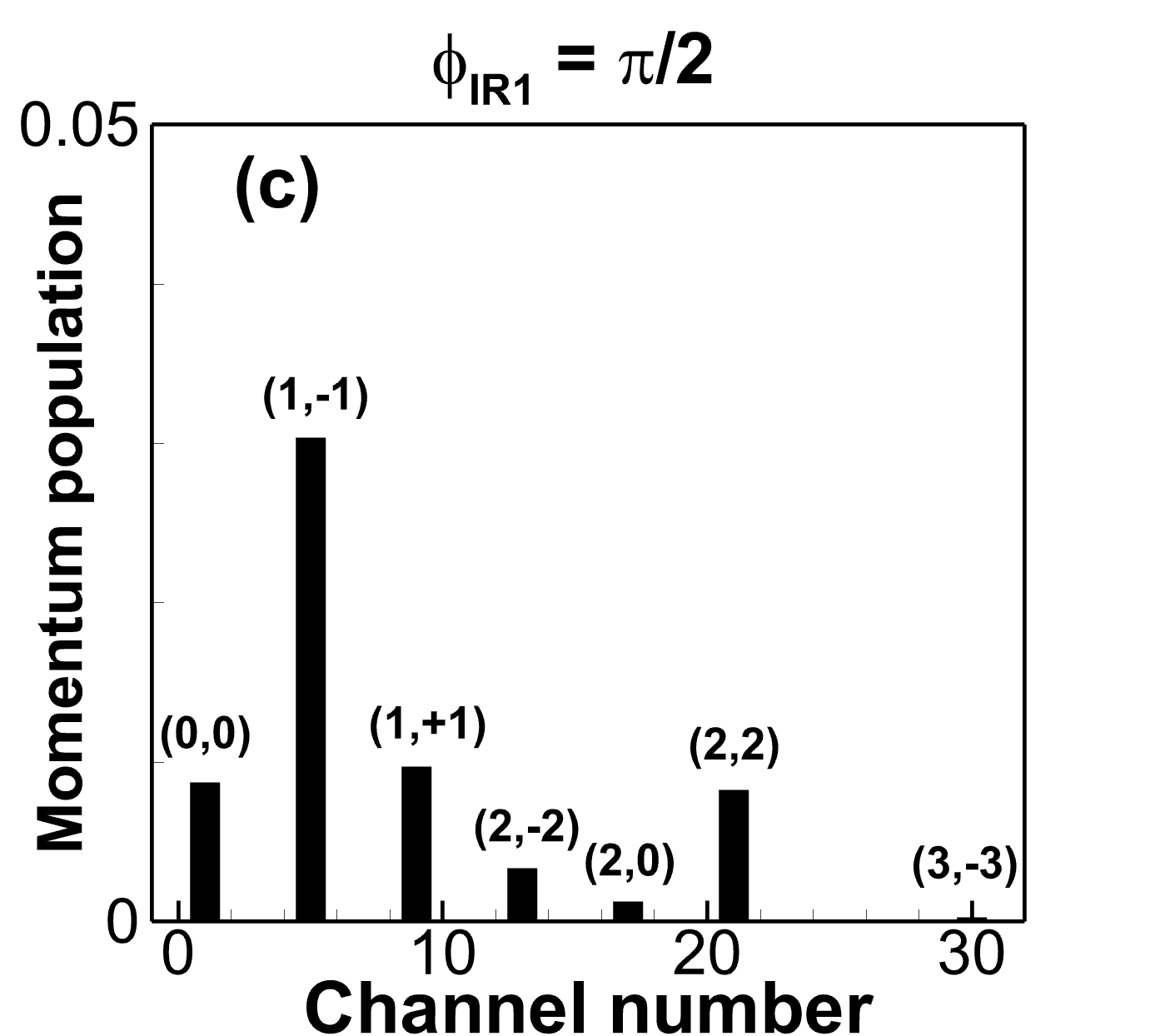}
  \includegraphics[width=0.49\linewidth]{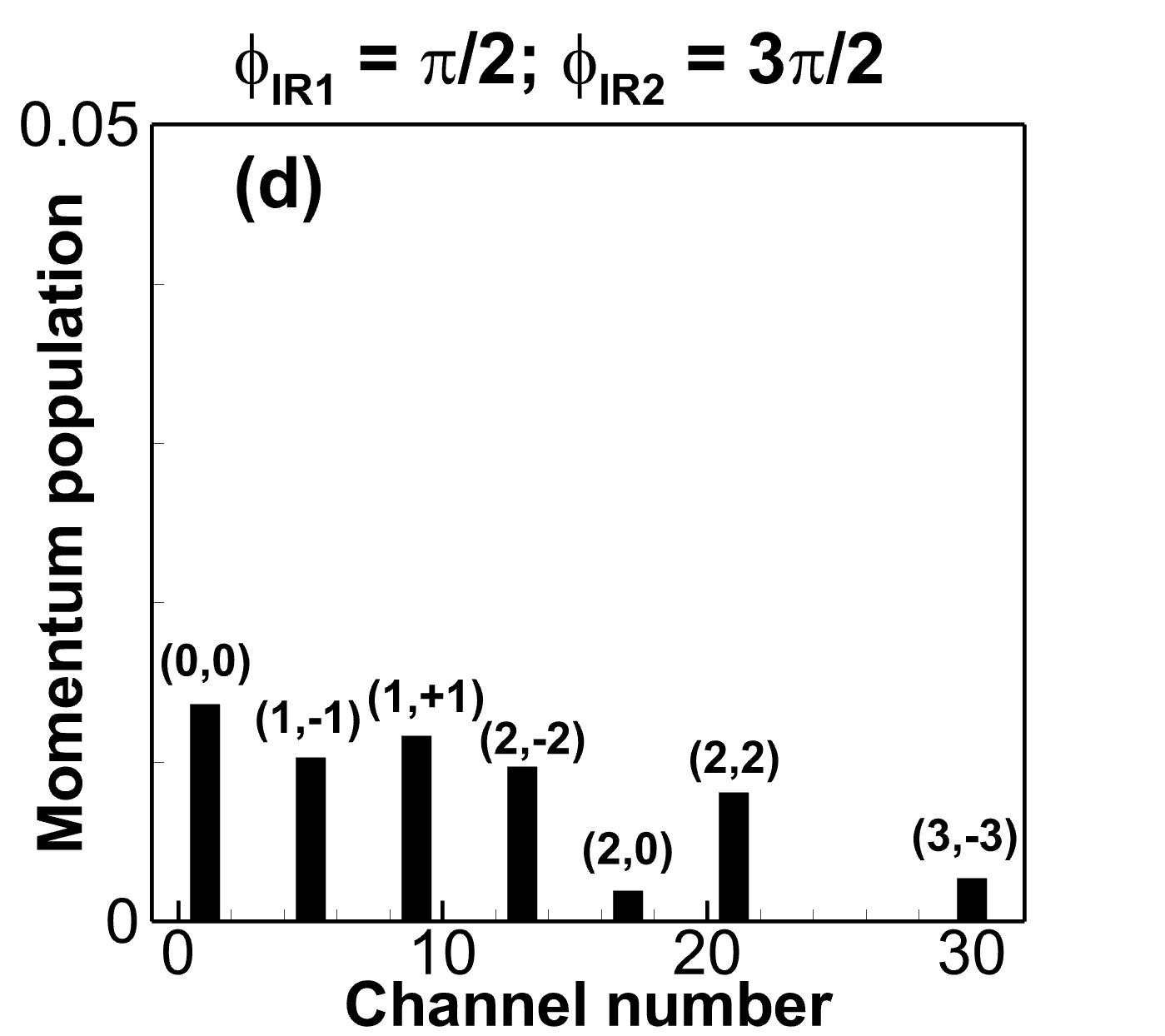} 
  \caption{Ionization channels populated by two CRCP XUV pulses in the (a)~absence or (b,c,d)~presence of IR field. Results in (b) and (c) are for one IR field with a CEP of $0$ and $\pi/2$; results in (d) are for two IR fields of CEPs of $\pi/2,3\pi/2$.}
  \label{Fig3}
  \vspace{-0.5cm}
\end{figure}

In the absence of the IR field, $\Phi_{\text{IR}}(\vp,\phi,t)=0$, and Eq.~\eqref{Amplitude} reduces to the first-order pertubation theory (PT) amplitude~\cite{Pronin2009,Ngoko2015} involving solely the two dominant ionization channels $(1,\pm1)$ in Fig.~\ref{Fig3}(a), where continuum electrons are described by Coulomb functions. The corresponding analytical TDP in this case reads:
\begin{equation}\label{TDP}
\mathcal{W}(\vp)=I_{\text{XUV},0}|A_0(p)|^2\sin^2(\theta)\cos^2(\Psi/2-\hat{\xi}\varphi),
\end{equation} 
where $\theta$ and $\varphi$ are the spherical angles of the photoelectron momentum vector $\vp$; the pulse helicity parameter $\hat{\xi}$ takes the value $+1$($-1$) for right-left (left-right) circularly polarized XUV pulses; and $A_0(p)\equiv \hat{F}(\epsilon)\Upsilon(p)$, where $\hat{F}(\epsilon)$ is the Fourier transform of the XUV pulse envelope $f_{\text{XUV}}(t)$ and $\Upsilon(p)$ is the radial matrix element between the ground and continuum states. As $p$ varies, $\hat{F}(\epsilon)$ exhibits a lump shape and $\Upsilon(p)$ monotonically decreasing with $p$, such that $|A_0(p)|^2$ has a bell shape. In the laser polarization plane ($\theta=\pi/2$) and for $\tau_{\text{XUV}}\ne0$, PT formula~\eqref{TDP} shows that the non-vanishing Ramsey phase part of $\Psi$ rotates the dipolar pattern (whose direction is dictated by the attopulse CEP difference $\phi_{12}$) to yield a spiral with two arms, transparently explaining the pattern in Fig.~\ref{Fig2}(a). The spiral equations are given by maxima (bright fringes) and zeros (dark fringes) of the kinematical factor in Eq.~\eqref{TDP}:
\begin{equation}\label{Spiral-equations}
\varphi^{\text{max},0}(E)=\hat{\xi}[n\pi+\Psi/2],
\end{equation} 
where $n$ is an integer (half integer) for maxima (zeros). Clearly, one sees that the spiral equations~\eqref{Spiral-equations} are those for Archimedean spirals type. For any photoelectron azimuthal angle $\varphi$, Eq.~\eqref{Spiral-equations} shows that successive dark or bright fringes are separated by $2\pi/\tau_{\text{XUV}}$, see for instance the PED shown in Fig.~\ref{Fig4} for forward emission ($\varphi=0$) or orthogonal emission ($\varphi=\pi/2$), which can be used to measure the time delay $\tau_{\text{XUV}}$ between the two XUV pulses. This observable shows the excellent contrast between dark and bright fringes separated by $\pi/\tau_{\text{XUV}}$.

Let us place one IR field at the first slit. Varying the IR-field intensity $I_0$ without changing its carrier frequency $\omega$, the obtained PMDs for any CEP do not differ from Fig.~\ref{Fig2}(a) whenever $I_0\leq 1$~GW/cm$^2$~\cite{SM}. This result, similar to that in the Feynman's thought experiment in space~\cite{Feynman1965}, indicates that some electrons get by without being disturbed or seen. Thus, when the light source gets dimmer we have interference, and the contrast between bright and dark fringes is again excellent as Eqs.~\eqref{TDP} and~\eqref{Spiral-equations} are still applicable. 

When $1\le I_0\le100$~GW/cm$^2$, the IR-field effect is only to impart a slight shift to the fringes without any raising of the dark fringes~\cite{SM}. Let us now crank up the IR-field intensity $I_0$ to $1$~TW/cm$^2$. With this increase of the rate at which the photons are emitted, it happens to be a photon around at the time the electron went through one of the slits. Depending on the IR-field waveform, we find that strong interaction of the IR field with either the first- or second-created electron wave packets gives rise to multiple ionization channels, as evidenced by Figs.~\ref{Fig3}(b,c). \textcolor{black}{Note that this scheme drastically differs from Ref.~\cite{Xiao2019}, where only the electron exiting the first femtoslit (with a width of $10.8$~fs) can interact with the infrared single-cycle $2.7$~fs light source placed at the middle of the two femtoslits, separated by $16.2$~fs.}  As the arguments of amplitudes of these channels introduce very different phases, mixing their contributions through phase averaging may cause a loss of contrast via dephasing visible in the PMD. The PMDs produced for an IR-field with a CEP of $0$ and $\pi/2$ are displayed in Fig.~\ref{Fig2}(b) and Fig.~\ref{Fig2}(c), respectively. In the former, while the excellent contrast is indicative of weak dephasing, the fringes are significantly shifted as can be seen in the PED shown in Fig.~S2(d) of~\cite{SM}, or by comparing the PMDs in Fig.~\ref{Fig2}(b) with Fig.~\ref{Fig2}(a). In the latter, the diffraction pattern along the IR-field polarization $+x$-axis is washed out especially at high photoelectron energies. For both CEPs there is an excellent contrast for perpendicular emission ($\varphi=\pm\pi/2$). These effects are better visualized in the PEDs in Fig.~\ref{Fig4}(a) for forward emission ($\varphi=0$) and in Fig.~\ref{Fig4}(b) for orthogonal emission. Changing the IR-field CEP from $\pi/2$ to $3\pi/2$ flips the directional loss of contrast in the PMD. 

To manipulate this effect over the entire range of photoelectron energy $E$, we place a second IR light at the second slit. Given the femto and atto timescales of the light sources and $\tau_{\text{XUV}}$, optical interference between the two IR fields takes place so that it is possible for the two electrons exiting the temporal slits to interact with the combined IR fields. This effect turns out to be very sensitive to the IR field waveforms. TDSE results shown in Fig.~S2(e) of Ref.~\cite{SM} for the CEP values of $\pi/2$ for both IR fields show no loss of contrast. On the contrary, for the IR-field CEPs of $\pi/2$ and $3\pi/2$, the PMD in Fig.~\ref{Fig2}(d) or the PED for forward emission in Fig.~\ref{Fig4}(c) now evidences a dramatic loss of contrast along the $\pm x$-axis. Over the entire range of photoelectron energy of interest, one observes a raising of all dark fringes, except for one which is possibly due to the difference in timescale between the light sources and widths of the temporal slits. Again, no loss of contrast is seen in the PMD [Fig.~\ref{Fig2}(d)] or the PED [Fig.~\ref{Fig4}(d)] for orthogonal emission along $\pm y$-axes.

To gain more insight into the interaction of the light source with an attoslit, we look at the ionization channel populations although information about the phases of those channels is not accessible. Focusing first on the case of a single light source, Fig.~\ref{Fig3}(b) for $\phi_{\text{IR}}=0$ shows that the channel $(1,-1)$ representing the first exiting electron is substantially diminished, while the channel $(1,+1)$ representing the second electron wave is unchanged. Meanwhile, Fig.~\ref{Fig3}(c) for $\phi_{\text{IR}}=\pi/2$ shows the inverse effect. For the case of two IR fields, Fig.~\ref{Fig3}(d) shows that the ionization channels $(1,\mp1)$ are both diminished with their contributions being now comparable. These interactions form a variety of interferometry in Fig.~\ref{Fig3}(b,c,d), where the appearance of channels $(2,0)$, $(0,0)$, and $(2,\pm2)$ by one-photon processes, or $(3,\pm3)$ by two-photon processes mediated by the IR field can be explained~\cite{SM,Bertolino2020}. In all cases, the dominance of the channel $(0,0)$ contribution over the channel $(2,0)$ contribution evidences the breakdown of Fano-propensity rules~\cite{Fano1985}, first evoked in~\cite{Busto2019}. 

\textcolor{black}{Besides the ionization channels $(1,\pm1)$, the other channels cannot be ruled out, evidencing the inapplicability of Eqs.~\eqref{TDP} and~\eqref{Spiral-equations} in describing the laser-assisted photoionization. We use standard SFA to interpret the TDSE results, where continuum electrons are described by Volkov states~\cite{Bertolino2020}. The amplitude~\eqref{Amplitude} can take the form,
\begin{equation}\label{SFA-amplitude}
\mathcal{A}=\sin\theta~\Upsilon(p)\sqrt{I_0} [\hat{G}^{(1)}(\epsilon) e^{i\varphi} + \hat{G}^{(2)}(\epsilon)e^{i(\Psi-\hat{\xi}\varphi)}],
\end{equation} 
where $\epsilon\equiv\varepsilon-\omega_{\text{XUV}}$, and $\hat{G}^{(1,2)}(\epsilon)=\text{FT}[f_{\text{XUV}}(t-\tau_{\text{IR}})e^{i\Phi_{\text{IR}}^{(1,2)}(\vp,\phi,t)}]$ is the Fourier transform of the XUV pulse envelope \textcolor{black}{weighted by the temporal gate $\Phi_{\text{IR}}^{(2)}(\vp,\phi,t)=\Phi_{\text{IR}}^{(1)}(\vp,\phi,t+\tau)$ used in the FROG~\cite{Quere2005,Mairesse2005} spectrogram for XUV pulse retrieval~\cite{Yakovlev2010} within the attosecond streaking technique~\cite{Drescher2001}. As the linearly polarized XUV pulse used in~\cite{Drescher2001} can be formed using two synchronous ($\tau_{\text{IR}}=0$) CRCP XUV pulses, moving these two ionizing pulses apart in time leads to the attosecond double-slit streak camera~\eqref{SFA-amplitude}, which suggests a dramatic angular dependence.} If the IR field is off, then $\Phi_{\text{IR}}(\vp,\phi,t)=0$ and $\hat{G}^{(1,2)}(\epsilon)=\hat{F}(\epsilon)$, meaning the TDP obtained using Eq.~\eqref{SFA-amplitude} reduces to PT formulas~\eqref{TDP}. When the IR field is on, the same physics applies for any CEP $\phi\equiv\phi_{\text{IR}}$ if electrons are emitted with $\vp$ along $\pm y$ axes, which are perpendicular to the polarization $x$-axis of $\vA_{\text{IR}}(t)$. For emission near the $x$ axis, Fig.~\ref{Fig3}(b) for $\phi=0$ suggests that $|\hat{G}^{(1)}(\epsilon)|\approx 0.4|\hat{G}^{(2)}(\epsilon)|$. Meanwhile, Fig.~\ref{Fig3}(c) for $\phi=\pi/2$ implies $|\hat{G}^{(2)}(\epsilon)|\approx 0.6|\hat{G}^{(1)}(\epsilon)|$. Thus, these two interferometries have nonuniform distributions. In the presence of two IR fields, not only the channels $(1,\pm1)$ contribute equally, other channels do as well. This leads to a uniform distribution, similar to the multimodal interferometric distribution in~\cite{Richard2025}, which led to decoherence due to quantum fluctuation of the ground state of phenylalanine. Regardless of the distribution, given that the amplitude for the isotropic channel $(0,0)$ is always large, it strongly mixes near the $x$-axis with the amplitude~\eqref{SFA-amplitude} including only the channels $(1,\pm1)$ with different phases and magnitudes. The non-uniform and uniform distributions are indicative of realization of the partial and almost total Feynman's thought experiment.} 

While a ruler formed by the interfering spiral patterns in the PMD was proposed~\cite{Xiao2019} to measure the \textcolor{black}{electron displacement $\vvr(t)=\int_{t}^{+\infty}\vA_{\text{IR}}(t')dt'$ induced by a short IR pulse; here this net excursion adds a nonstationary position-dependent phase $\Phi_{\text{IR}}(\vp,\phi,t)\approx \vp\cdot \vvr(t)$ to the electron wave}, which is thought to be responsible for the loss of contrast in the PMD via dephasing processes, \textcolor{black}{not by decoherence. This interpretation is consistent with previous works. Indeed, an example for dephasing was given in~\cite{Rui-Feng2014} where a rotating ground glass disk placed just before the slits adds a time-dependent phase to the light wave, or in~\cite{Chen2018} where a smooth random potential distribution placed just after the slits adds a position-dependent phase to the electron wave.}

In summary, in the scheme where ionization processes in matter driven by CRCP attopulses produce spiral patterns in the PMD with excellent contrast between dark and bright Ramsey interference~\cite{Ramsey1950} fringes, we have realized the Feynman thought experiment by showing that a femtolight source placed at the locations of a slit can strongly interact with the electrons as they exit the attosecond temporal slits. Being very sensitive to the light source waveform, this interaction can destroy the interference fringes (leading to a loss of contrast) despite the difference in timescale between the femtosecond light source and attosecond width of the slits. \textcolor{black}{Not shown here, the physics discussed has been conclusively tested for single-photon or multiphoton ionization processes from atoms (H and He) or aligned molecules (H$_2$).}  Our results for a single point in the interferogram of the proposed attosecond double-slit streak camera suggest its potential application in attosecond metrology. \textcolor{black}{This study on contrast loss is exciting in view of the need to understand and control the detrimental experimental effect of contrast loss and for fundamental studies on the transition from the classical to the quantum regime~\cite{Chen2018}.}


\section*{Acknowledgements}

Research is supported by the US Department of Energy~(DOE),
Office of Science, Basic Energy Sciences~(BES), under Award No.~DE-SC0021054. Computations were done using Swan of the Holland Computing Center of the University of Nebraska, which receives support from the Nebraska Research Initiative.

\end{document}